\documentclass{emulateapj}
\usepackage{epsfig}

\newcommand{\etal}{et al.~}
\def\gsim{\lower 2pt \hbox{$\, \buildrel {\scriptstyle >}\over
{\scriptstyle \sim}\,$}}
\def\lsim{\lower 2pt \hbox{$\, \buildrel {\scriptstyle <}\over
{\scriptstyle \sim}\,$}}

\def\HI{H~{\scriptsize I}}
\def\hii{H~{\scriptsize II}}

\def\nv{N~{\scriptsize V}}

\def\civ{C~{\scriptsize IV}}

\def\oviii{O~{\scriptsize VIII}}
\def\ovii{O~{\scriptsize VII}}
\def\ovi{O~{\scriptsize VI}}
\def\oi{O~{\scriptsize I}}
\def\oii{O~{\scriptsize II}}

\def\oiii{O~{\scriptsize III}}
\def\nei{Ne~{\scriptsize I}}
\def\neii{Ne~{\scriptsize II}}
\def\neiii{Ne~{\scriptsize III}}
\def\neviii{Ne~{\scriptsize VIII}}
\def\neix{Ne~{\scriptsize IX}}
\def\nex{Ne~{\scriptsize X}}

\def\mgii{Mg~{\scriptsize II}}

\def\mgxi{Mg~{\scriptsize XI}}

\def\fexvii{Fe~{\scriptsize XVII}}

\shortauthors{Yao \etal}
\shorttitle{The Multiphase ISM toward Cyg X-2}

\begin{document}
\slugcomment{\em Accepted for publication in the  Astrophysical Journal}
\title{High resolution X-ray spectroscopy of the multiphase interstellar 
	medium toward Cyg X-2}
\author{Yangsen Yao\altaffilmark{1},
        Norbert S. Schulz\altaffilmark{2},
        Ming F. Gu\altaffilmark{3},
        Michael A. Nowak\altaffilmark{2}, and 
        Claude R. Canizares\altaffilmark{2}}

\altaffiltext{1}{University of Colorado, CASA, 389 UCB, Boulder, CO 80309;
 yaoys@colorado.edu}
\altaffiltext{2}{Massachusetts Institute of Technology (MIT) Kavli Institute
 for Astrophysics and Space Research, 70 Vassar Street, Cambridge, MA 02139}
\altaffiltext{3}{Lawrence Livermore National Laboratory, 7000 East Avenue, 
Livermore, CA 94550}

\begin{abstract}

High resolution X-ray absorption spectroscopy is a powerful diagnostic tool 
for probing chemical and physical properties of the interstellar medium (ISM) 
at various phases. 
We present detections of K transition absorption lines 
from the low ionization ions of \oi, \oii, \nei, \neii, and \neiii, and the
high ionization ones of \ovi, \ovii, \oviii, \neix, and \mgxi, as well as 
details of neutral absorption edges from Mg, Ne, and O
in an unprecedented high quality spectrum of the 
low mass X-ray binary Cyg~X-2. These absorption features trace the intervening
interstellar medium which is indicated by the unshifted line centroids with 
respect to the rest frame wavelengths of the corresponding atomic transitions. 
We have measured the column densities of each ion.
We complement these measurements with the radio
\HI\ and optical H$\alpha$ observations toward the same sight line 
and estimate the mean abundances of Ne, O, and Mg in the cool phase to
Ne/H=$0.84^{+0.13}_{-0.10}\times10^{-4}$,
O/H=$3.83^{+0.48}_{-0.43}\times10^{-4}$, and
Mg/H=$0.35^{+0.09}_{-0.11}\times10^{-4}$, 
and O and Mg in the hot phase to O/H=$5.81^{+1.30}_{-1.34}\times10^{-4}$ and
Mg/H=$0.33^{+0.09}_{-0.09}\times10^{-4}$, respectively.
These results indicate a mild depletion of oxygen into dust grains 
in the cool phase and little or no depletion of magnesium.
We also find that absorption from highly ionized ions in the hot Galactic 
disk gas can account for most of the absorption observed toward the 
extragalactic sight lines like Mrk~421. The bulk of the observed \ovi\
likely originates from the conductive interfaces between the cool and hot
gases, from which a significant amount of \nv\ and \civ\ emission
is predicted.

\end{abstract}


\section{Introduction }
\label{sec:intro}

It is generally believed that the Galactic diffuse interstellar medium 
(ISM) has three major phases to a large extent regulated by supernova explosions (SNe): 
cold/warm neutral, cold/warm ionized, and hot ionized 
(e.g., \citealt{mck77, fer01}).
The neutral phase, as traced by Ly$\alpha$ and 21 cm emission 
of \HI, is mainly concentrated in the Galactic plane within a  
height of no more than several hundred parsec 
(pc, e.g., Lockman \etal 1986). The distribution
of the diffuse warm ionized gas, which is traced by H$\alpha$ emission and 
pulsar dispersion measures, can be characterized as an exponential decay with 
a vertical scale height of $\sim1$ kpc 
(e.g., Berkhuijsen \etal 2006). The hot gas at temperatures of $\sim10^{5-7}$ 
K is presumably sustained by SNe and massive stellar winds and is
traced by highly ionized absorption lines and soft X-ray background emission. 
It can extend vertically as far as several kpc from the Galactic plane
(e.g. \citealt{sav90, sno97, yao05}).

The chemical and physical properties of interstellar media in the various 
phases carry important information about formation and evolution of galaxies.
Stars are formed out of the ISM, and metals are all produced in stars and 
released back to the ISM through various forms of stellar feedback. It is 
believed that massive 
galaxies like our own are still evolving by accreting intergalactic 
matter to replenish the fuel for star formation activity. The accreted
material, which is presumably primordial or metal poor, 
is further altering the composition in the ISM phases.
Therefore, the similarities and dissimilarities of the 
chemical abundances between stars and the ISM provide
us with information about the interplay among the
different phases of the ISM as well as the star formation and evolution 
history. Characterizing the spatial distribution of the hot 
ISM in spiral galaxies  
is also a vital step to distinguish the small scale ($<10$ kpc)
disk component from the large scale ($>$20 kpc) halo
hot gas; the latter is presumably a reservoir of the shock-heated in-falling
gas during the galaxy formation/evolution and is believed to provide one 
possible solution to the ``missing baryon'' problem for an individual dark 
matter halo (\S~\ref{sec:halo};
e.g., Sommer-Larsen 2006 and references therein; Yao \etal 2008).

X-ray spectroscopy can provide a powerful diagnostic of the
physical and chemical properties of the multiphase ISM, not only because the 
K-shell transitions of carbon to iron and the L-shell transitions of silicon to
iron are accessible in the X-ray band (e.g., \citealt{bar79, york83, pae03}),
but also because 
the multiphase ISM can be probed by a single abundant metal element
at different charge states (e.g., Yao \& Wang 2006; Juett \etal 2006).
The latter is particularly striking, and can be utilized to infer 
ionization states and thus thermal properties of the ISM without any
confusion of the relative abundances between different elements. 

X-rays interact with the ISM via scattering and absorption. Scattered 
photons by dust grains are detected as a diffuse X-ray halo around the 
emitting source which carries information of chemical compositions, the
size and the spatial distributions of dust grains, as well as the 
distance of the X-ray source (e.g., 
\citealt{ove65, pre96, pre00, xiang05, xiang07}).
Although a detailed study of dust 
scattering is beyond the scope of this work, its importance 
towards inferring the chemical abundances of the ISM has been recognized 
(e.g., Juett \etal 2006). X-rays absorbed by the ISM imprint absorption
lines and/or edges in the X-ray spectrum of the background source;
the ISM at different phases produces absorption features for different
charge states. These features can be used to determine the
average chemical abundances and ionization states of the responsible elements
in the ISM. 

\citet{sch86} first demonstrated the potential of high resolution 
X-ray spectroscopy for studies of the ISM by
presenting a prominent oxygen K absorption edge and a marginal narrow atomic 
oxygen $1s-2p$ absorption transition at the expected wavelength in a Crab 
spectrum. However, because of the limited sensitivity of the previous X-ray 
instruments, {\sl Einstein observatory} for example, 
$E/\Delta E\sim50-200$ across the 
wavelength of 7--46 \AA, the power of the X-ray spectroscopy has not been 
extensively explored until recently.

The launches of {\sl Chandra} and {\sl XMM-Newton} opened a new
era in the study of the ISM via high resolution X-ray spectroscopy. 
The diffraction grating instruments aboard these two observatories offer a 
resolving power of $E/\Delta E \ge 200$ across most of the soft X-ray 
band, which, for the first time allows us to resolve the absorption features
in great detail and enables us to systematically study the multiple phase
ISM via detections of multiple absorption edges/lines of ions at different 
ionization states \citep{pae01, sch02, page03, jue04, yao05}.
In addition, these high spectral resolution spectrometers, especially the
high energy transmission grating spectrometer (HETGS; Canizares \etal 2005) 
aboard {\sl Chandra} with a spectral resolution of $\sim1,000$, have
pushed the detections to the systematic uncertainties of atomic databases.
For instance, in a systematic study of the O K absorption structures, 
\citet{jue04} found that the centroid 
wavelength of the interstellar \oi\ 1s--1p line is $\sim30-50$ m\AA\ offset
from that predicted from the theoretical calculations. Clearly, accurate
measurements of interstellar absorption features, 
which are supposed to be at the rest frame wavelengths,
will provide valuable references for theoretical calculations and 
laboratory measurements.

In this work, we report high-quality detections of multiple absorption
lines and edges in a grating spectrum of the low mass X-ray binary (LMXB)
Cyg~X-2 obtained with the HETGS. The source is well suited as 
a background source to study ISM absorption. Its  
galactic coordinates $l,b=87\fdg33, -11\fdg32$ and distance 
$D\sim7-12$ kpc mean that it is located $\sim1.4-2.4$ kpc above the 
Galactic plane \citep{cow79,mcc84, sma98, oro99}.
Therefore its pathlength samples a bulk of the Galactic disk gas with very 
little confusion from
the central region of the Galaxy where the diffuse emission is greatly 
enhanced. The radial velocity of its optical companion suggests
a systematic velocity about $-220~{\rm km~s^{-1}}$ \citep{cow79, cas98} of
the binary system, which  
is essential for distinguishing any absorptions produced in the circumstellar
material of the source from that in the ISM (\S~\ref{sec:origin}). 

The paper is organized as follows. In \S~\ref{sec:atomic}, 
we list the atomic data we are referencing and also briefly introduce the 
absorption line model utilized in this work. 
In \S~\ref{sec:observation}, we describe the observations, 
data reduction, line identifications, as well as measure equivalent widths (EW)
and optical depths of absorption features.
In \S~\ref{sec:results}, we determine the ionic column densities
and discuss the properties of the ISM in the cold/warm (\S~\ref{sec:CISM}) and 
hot phases (\S~\ref{sec:HISM}) separately. We then discuss the location
of these absorption features (\S~\ref{sec:origin}), estimate the systematic
uncertainties of their centroid wavelengths (\S~\ref{sec:restframe}), explore 
the mystery of the undetected \ovii\ K$\alpha$ line (\S~\ref{sec:oviimis}), 
compare the chemical abundances in different phases (\S~\ref{sec:chemical}), 
and estimate the contribution of the Galactic disk hot gas to the absorption
observed toward the extragalactic sources (\S~\ref{sec:halo}).
Final results are summarized in \S~\ref{sec:sum}.

Throughout this paper, we reference the solar chemical abundances from 
\citet{and89}, and quote the uncertainty of a single floating parameter
at 1$\sigma$ significance levels unless otherwise specified. 
We also assume that collisional ionization equilibrium 
(CIE; Sutherland \& Dopita 1993) is reached in the hot phase ISM.

\section{Atomic data and an absorption line model}
\label{sec:atomic}

\citet{ver96} and NIST 
\footnote{http://physics.nist.gov/PhysRefData/ASD/lines\_form.html}
are two commonly referenced databases for atomic lines in the X-ray.
However, neither of the two has included the K transitions of 
the neutral and mildly ionized metal elements like oxygen and neon,
from which the interstellar lines have now been commonly observed 
(e.g., Juett \etal 2004, 2006). 
Recently, several groups have calculated and updated these databases
(e.g., Behar \& Netzer 2002; Garcia \etal 2005), but the atomic data for some
transitions are still missing and/or not available in proper formats
(Table~\ref{tab:lineparameters}). To this regard, we used the 
Flexible Atomic Code (FAC 
\footnote{FAC is available at http://kipac-tree.stanford.edu/fac/}
; Gu 2003) to calculate the K transitions of all 
charge states for oxygen, neon, and magnesium. For He-like
ions, we calculated transitions from the $1s-2p$ to $1s-7p$, 
whereas for H-like ones we calculated up to the $1s-12p$.
For all the lines detected or used in this work (\S~\ref{sec:observation}),
Table~\ref{tab:lineparameters} lists the basic parameters of the transitions,
which include the line centroid wavelength ($\lambda$), 
transition oscillation strength ($f_{ij}$), and 
the natural width ($\gamma$; including radiative and Auger decay rates).
For those transitions with a cluster of lines that
can not be resolved with the HETGS resolution, we take a summation of
their $f_{ij}$ and average the $\lambda$ and the $\gamma$ values
with respect to the corresponding $f_{ij}$ values. For a comparison, we also
list the parameters from several other references when they are available. 
The measured line centroids in this work are also listed in
the last column (Table~\ref{tab:lineparameters}).
It is important to point out that these parameters can also be extracted
from XSTAR 
\footnote{http://heasarc.gsfc.nasa.gov/docs/software/xstar/xstar.html}
\citep{kal04} and APED 
\footnote{http://cxc.harvard.edu/atomdb/sources\_aped.html}
\citep{smi01} with some calculations.

\begin{deluxetable*}{lc|ccc|ccc|ccc|c}
  \tablewidth{0pt}
  \tablecaption{Line parameters}
  \tablehead{
    &            &  \multicolumn{3}{c}{This work} & \multicolumn{3}{c}{NIST$^a$} & \multicolumn{3}{c}{Verner \etal$^b$} & Detection$^c$ \\
    &            &    $\lambda$ &           & $\gamma$ & $\lambda$ &           & $\gamma$ & $\lambda$ &           & $\gamma$ & $\lambda$\\
Ion & Transition &    (\AA)      & $f_{ij}$  & ($10^{12}$ s$^{-1}$) &  (\AA)      & $f_{ij}$  & ($10^{12}$ s$^{-1}$) & (\AA)      & $f_{ij}$  & ($10^{12}$ s$^{-1}$) & (\AA) } 
  \startdata
\oi\     & 1s--2p  & 23.4760  & 0.113 & 163  &   23.4475  & 0.191    & 248        & 23.532   & $\cdots$ & $\cdots$ &  $23.508^{+1.6}_{-1.6}$ \\ %
\oi\     & 1s--3p  & 22.8982  & 0.006 & 113  &   $\cdots$ & $\cdots$ & $\cdots$   & 22.907   & $\cdots$ & $\cdots$ &   	   $\cdots$ \\ 
\oii\    & 1s--2p  & 23.3119  & 0.192 & 151  &   23.310   & 0.212    & 204        & $\cdots$ & $\cdots$ & $\cdots$ &  $23.348^{+4.2}_{-4.2}$ \\ %
\nei\    & 1s--3p  & 14.3201  & 0.014 & 220  &   $\cdots$ & $\cdots$ & $\cdots$   & 14.298   & $\cdots$ & $\cdots$ &  $14.294^{+1.5}_{-1.3}$ \\ %
\neii\   & 1s--2p  & 14.5990  & 0.067 & 290  &   16.431   & 0.062    & 5.80       & 14.605   & $\cdots$ & $\cdots$ &  $14.605^{+1.0}_{-1.0}$ \\ %
\neii\   & 1s--3p  & 14.0070  & 0.025 & 211  &   $\cdots$ & $\cdots$ & $\cdots$   & $\cdots$ & $\cdots$ & $\cdots$ &  $14.001^{+2.0}_{-1.2}$ \\ %
\neiii\  & 1s--2p  & 14.4990  & 0.146 & 276  &   14.526   & 0.106    & 255        & 14.518   & $\cdots$ & $\cdots$ &  $14.507^{+2.0}_{-2.1}$ \\ %
\neiii\  & 1s--3p  & 13.6977  & 0.029 & 191  &   $\cdots$ & $\cdots$ & $\cdots$   & $\cdots$ & $\cdots$ & $\cdots$ &  $13.690^{+6.3}_{-1.5}$ \\ %
\hline
\ovi\    & 1s--2p  & 22.0403  & 0.592 & 9.52 &   $\cdots$ & $\cdots$ & $\cdots$   & $\cdots$ & $\cdots$ & $\cdots$ &  $22.026^{+4.0}_{-4.0}$ \\ %
\ovii\   & 1s--2p  & 21.6020  & 0.731 & 3.48 &   21.6020  & 0.695    & 3.31       & 21.6019  & 0.696    & 3.32     &   	    $\cdots$         \\ %
\ovii\   & 1s--3p  & 18.6538  & 0.148 & 0.94 &   18.6270  & 0.146    & 0.94       & 18.6288  & 0.146    & 0.94     &  $18.625^{+2.6}_{-2.5}$ \\ %
\ovii\   & 1s--4p  & 17.7999  & 0.055 & 0.39 &   $\cdots$ & $\cdots$ & $\cdots$   & 17.7686  & 0.055    & 0.39     &   	    $\cdots$         \\ %
\oviii\  & 1s--2p  & 18.9670  & 0.277 & 2.57 &   18.9671  & 0.277    & 2.57       & 18.9689  & 0.832    & 2.57     &  $18.964^{+2.0}_{-1.7}$ \\ %
\oviii\  & 1s--3p  & 16.0058  & 0.079 & 0.68 &   16.0059  & 0.079    & 0.69       & 16.0059  & 0.158    & 0.69     &  $16.003^{+6.7}_{-6.7}$ \\ %
\neviii\ & 1s--2p  & 13.6603  & 0.628 & 14.1 &   13.6550  & $\cdots$ & $\cdots$   & $\cdots$ & $\cdots$ & $\cdots$ &   	   $\cdots$          \\ %
\neix\   & 1s--2p  & 13.4497  & 0.751 & 9.23 &   13.4470  &  0.721   & 8.87       & 13.4471  & 0.724    & 8.90     &  $13.445^{+1.1}_{-1.2}$ \\ %
\neix\   & 1s--3p  & 11.5568  & 0.150 & 2.49 &   11.5470  &  0.149   & 2.48       & 11.5466  & 0.149    & 2.48     &  $11.549^{+1.4}_{-3.4}$ \\ %
\nex\    & 1s--2p  & 12.1337  & 0.415 & 6.27 &   $\cdots$ & $\cdots$ & $\cdots$   & 12.1339  & 0.832    & 6.28     &   	     $\cdots$        \\ %
\mgxi\   & 1s--2p  & 9.1699   & 0.764 & 20.2 &   9.1689   &  0.741   & 19.6       & 9.1688   & 0.742    & 19.6     &  $ 9.170^{+0.6}_{-1.2}$ \\ %
\fexvii\ & 2p--3d  & 15.0150  & 2.950 & 29.1 &   15.0150  & 2.31     & 22.8       & 15.0150  & 2.950    & 29.1     &   	     $\cdots$
\enddata
\label{tab:lineparameters} 
\tablecomments{Horizontal dots indicate that the values are unavailable.
	$^a$ Values are listed in these columns for \oi-\oii\ and 
	\nei-\neiii\ are from \citet{gar05} and \citet{beh02}, respectively.
	$^b$ Values are listed in these columns for \oi-\oii\ and 
	\nei-\neiii\ are from \citet{gor00} and \citet{gor05}.
	$^c$ Wavelengths are detected in this work, and the uncertainties are
	in units of m\AA.}
\end{deluxetable*}

For an absorption line produced in an intervening gas, the radiative transfer
at different wavelengths can be characterized with these parameters
($\lambda$, $f_{\rm ij}$, and $\gamma$) together with the 
physical properties of the absorbing gas (i.e., total ionic column 
density $N_i$ and its velocity dispersion $v_b$). 
The absorption optical depth is linearly proportional 
to $f_{\rm ij}$, and the absorption line profile, which is a convolution of 
the intrinsic Lorenz profile (depending on $\lambda$, $\gamma$, and $N_i$)
with the Doppler broadening (characterized by $v_b$), can be approximated 
as a Voigt function. For a hot absorbing gas at CIE, 
$N_i$ is function of the gas temperature \citep{sut93}.
Therefore, multiple absorption lines
of different elements at various charge states can provide a diagnostic
of the thermal and chemical properties of the gas (e.g., Yao \& Wang 2005).
Following the description of the radiative transfer process by \citet{ryb79},
\citet{yao05} developed an absorption line model, {\sl absline}, which takes
$\lambda$, $f_{\rm ij}$,  $\gamma$, $N_i$, and $v_b$ as input parameters for 
each transition. This model, compared to the {\it additive} Gaussian model
as commonly adopted in modeling absorption lines, is a {\it multiplicative}
model that can take into account of the line saturation automatically.
It can also be used to fit a single absorption line as well as to jointly
analyze multiple absorption lines (see Yao \& Wang 2005 for details). 
In this work, we utilize this line model to fit each observed 
absorption line and also jointly analyze the highly ionized ones to
infer the ionic column densities and the thermal and chemical properties
of the hot ISM (\S~\ref{sec:results}).

\section{Observations and Data reduction}
\label{sec:observation}

{\sl Chandra} observed Cyg~X-2 with grating instruments six times, twice with 
the low energy transmission grating (LETG) and four times with the HETG 
(Table~\ref{tab:observation}). \citet{tak02} analyzed the neutral and low
ionization O and Ne absorption features detected in the longer LETG 
observation and derived the O and Ne abundances 
(see \S\S~\ref{sec:CISM} and \ref{sec:chemical} for more discussions). 
In this work, we focus on the four HETG observations, among which the first
two (ObsIDs 1016 and 1102) were performed with the 
regular timed-exposure (TE) observation mode while the last two were
with the continuous-clocking (CC) mode. 

\begin{deluxetable}{llcl}
  \tablewidth{0pt}
  \tablecaption{{\sl Chandra} Grating Observation log of Cyg~X-2}
  \tablehead{
         &                   & Exposure \\
ObsID    & Observation Date  &  (ks)  & grating}  
  \startdata
87   & 2000 Apr. 24 & 30 & LETG-HRC  \\
111  & 1999 Nov. 11 & 3  & LETG-ACIS \\
1016 & 2001 Aug. 12 & 15 & HETG-ACIS \\
1102 & 1999 Sep. 23 & 29 & HETG-ACIS \\
8599 & 2007 Aug. 23 & 71 & HETG-ACIS \\
8170 & 2007 Aug. 25 & 79 & HETG-ACIS 
\enddata
\label{tab:observation} 
\end{deluxetable}

We calibrated all observations using CIAO 4.0
and CALDB 3.4.5 \footnote{http://asc.harvard.edu/ciao/}. 
For a grating observation, the first and crucial step is to determine the 
zeroth order source position in order to fix the wavelength scale. Since 
Cyg~X-2 is very bright, the saturated detections produced a piled-up hole 
in its zeroth order image for the observations 
taken with the TE mode. To avoid any 
confusion in searching for the source position with the
standard script, we used the cross between the read-out streak and the medium 
energy grating (MEG) arm as the source position  
\footnote{This procedure is similar to the algorithm adopted by the 
script {\sl findzo}, http://space.mit.edu/cxc/analysis/findzo/index.html}.
The other steps, 
including extracting the spectrum, building the energy redistribution matrix 
file (RMF), and calculating the ancillary response file (ARF), followed the
standard procedures 
\footnote{http://asc.harvard.edu/ciao/threads/gspec.html}.
For each observation, we combined the spectra of the negative and the positive
grating arms, and then followed the procedure described in \citet{yao06}
co-adding the spectra and the RMFs and ARFs of all four 
observations. Since we are only interested in
ISM absorption features which are expected to appear at longer ($>8$ \AA)
wavelengths, we exclusively focused on the range of 9.0--24.0 \AA\ in the 
first order MEG spectrum. At these long wavelengths we do not expect 
calibration problems stemming from our use of the CC mode 
(see \citealt{sch09}). We conducted our data analysis using the software
XSPEC (version 11.3.2).

We adopted an absorbed power-law model $varabs(powerlaw)$
to fit the continuum.  
To avoid the possible effects of the ISM absorption lines on the continuum
determination, we grouped the
spectrum to a 0.1 \AA\ bin size throughout the spectral range. To characterize
the absorption of O, Ne, and Mg, the three most abundant elements whose
K absorption edges are in the MEG wavelength range, we set their abundances to
zero in $varabs$ model and added three absorption $edge$ models at the 
corresponding energies. During the continuum fit, we also manually inserted 
several very broad Gaussian (with $\sigma>10^3~{\rm km~s^{-1}}$)
profiles in addition to the power law to correct for very broad 
bumps and wiggles in the spectrum.
Since the ISM absorption lines are expected to be very narrow 
($\sigma<350~{\rm km~s^{-1}}$; e.g., \citealt{yao05,yao06}), their 
measurements depend on the relative flux ratio between the local continuum
and the absorption dip. Likewise, the absorption edge measurements depend
on the relative flux before and after the edge. The application of broad
Gaussian profiles to smooth out the continuum therefore does not affect our 
absorption feature measurements. After obtaining the best fit to the 
continuum, we restored the spectral resolution and used a 10 m\AA\ spectral
bin measuring and analyzing the narrow absorption lines and edges. 
Table~\ref{tab:absedge} lists the three K absorption edge wavelengths, optical
depth, and errors.

\begin{deluxetable}{lccccc}
  \tablewidth{0pt}
  \tablecaption{Absorption Edge Parameters and measurements}
  \tablehead{
        &  cross section$^a$     & $\lambda^a$   & $\lambda^b$ & $\tau^b$ \\
	& ($10^{-19}~{\rm cm^{-2}}$) & (\AA) & (\AA) & ($10^{-2}$)} 
  \startdata
O       & 5.2718  & 23.0452 & $22.808^{+7.3}_{-7.1}$ & $36.1^{+2.7}_{-2.6}$ \\
Ne      & 3.6696  & 14.3003 & $14.279^{+3.9}_{-5.0}$ & $6.5^{+0.6}_{-0.6}$ \\
Mg      & 2.3857  &  9.5124 & $9.491^{+2.8}_{-3.7}$  & $2.0^{+0.3}_{-0.3}$
\enddata
\label{tab:absedge} 
\tablecomments{Wavelength uncertainties are given at 1$\sigma$ level and 
	in units of m\AA. 
	$^a$ Values are adopted from \citet{bal92}. 
	$^b$ Values are measured in this work.}
\end{deluxetable}

\begin{figure*}
   \plotone{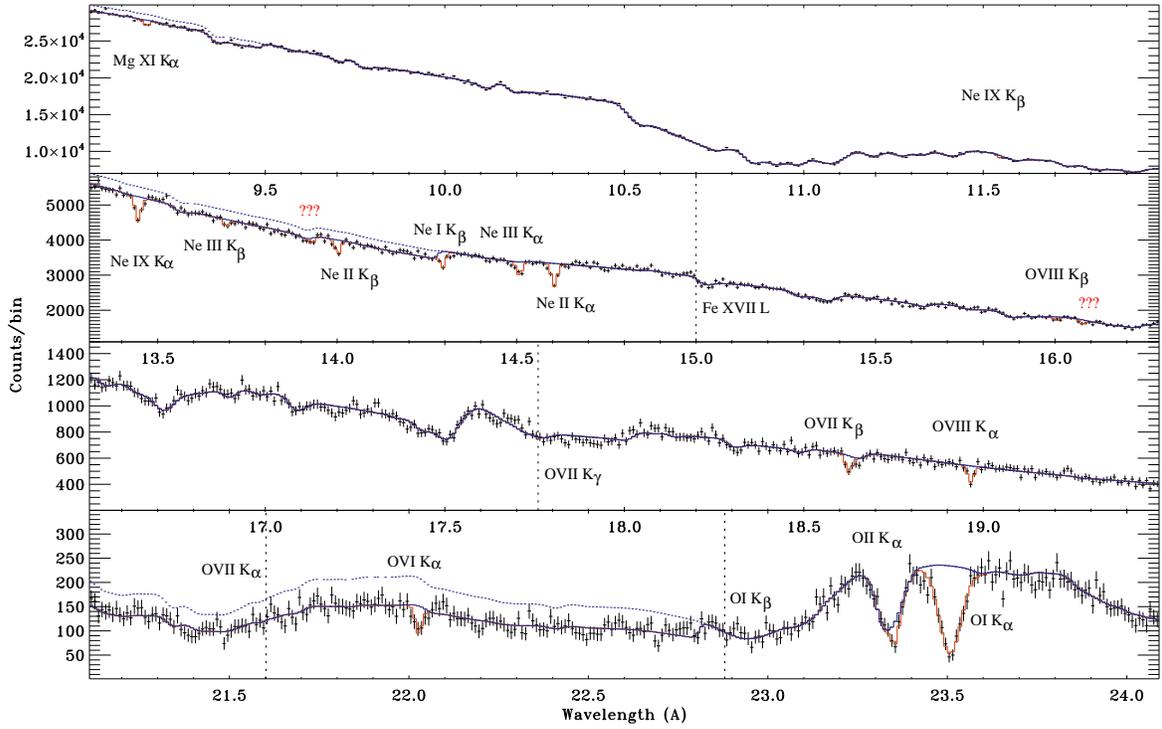}
   \caption{Co-added MEG spectrum with the best fit continuum (thick blue 
	lines). The dotted lines mark the continuum levels without
	the Mg, Ne, and O edge absorptions. Red histograms mark the observed 
	absorption lines, and question marks indicate those unidentified ones.
	The vertical dotted lines mark those undetected lines, which are used
	to constrain our diagnostic in this work. The bin-size is 10 m\AA. 
    \label{fig:counts}
  }
\end{figure*}

\begin{figure*}
   \plotone{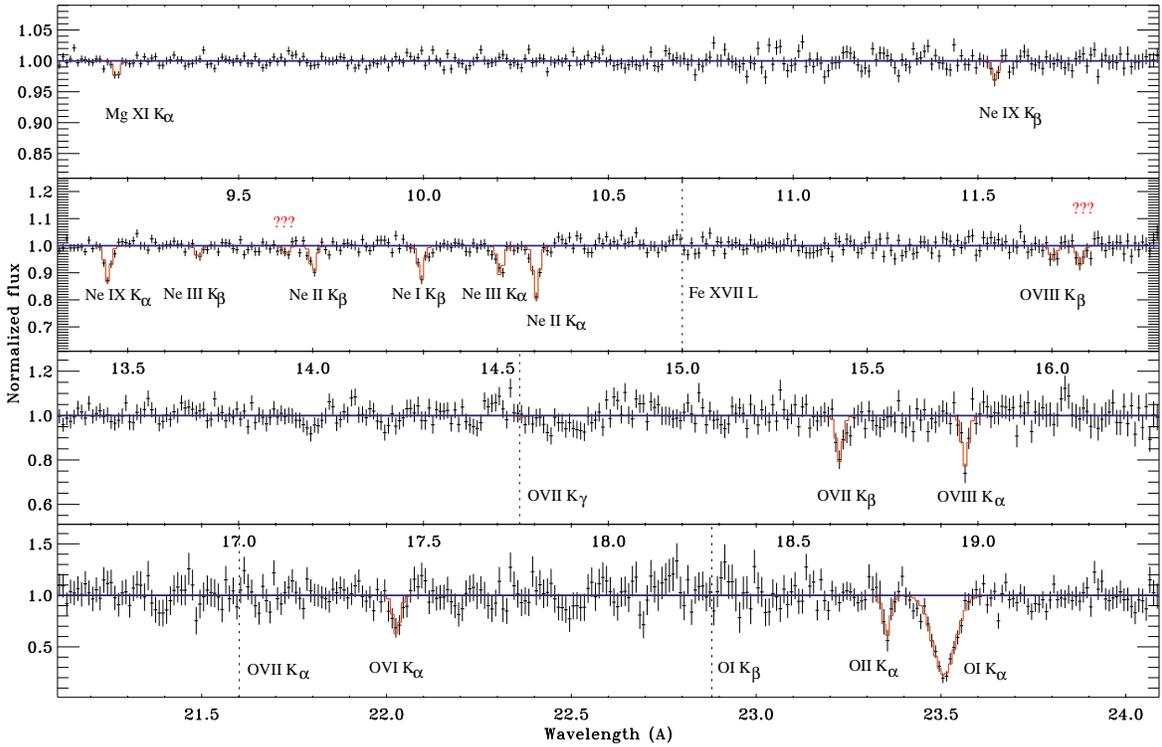}
   \caption{Same as the Figure~\ref{fig:counts}, but spectrum is normalized to the
	best fit continuum.
    \label{fig:ratio}
  }
\end{figure*}

We then searched for and measured absorption lines throughout the spectrum.
We inserted a negative Gaussian profile with the $\sigma$ fixed to 
$50~{\rm km~s^{-1}}$ and scanned the entire spectral range with a step
of 20 m\AA.  We recorded the Gaussian centroid wavelengths in cases when
the fits were improved by $\Delta\chi^2>5$, which corresponds a false 
detection possibility of $<2.5\%$ for a continuous feature. The successive 
records were regarded as a single detection. We recorded 17 detections and
found the one at 17.201 \AA\ is at exact position as one of the Fe L edges
\citep{jue06}. The study of the Fe L edges is beyond of the scope of this 
work, and the feature at 17.201 \AA\ will not be discussed any further.
Among the other 16 detections, 14 lines were easily identified by 
comparing their centroids with the rest frame wavelengths from the 
references listed in Table~\ref{tab:lineparameters}, and the remaining
2, with relatively lower significances (Figs.~\ref{fig:counts} and
\ref{fig:ratio}), were unidentified and could be false detections. 
We then used 16 Gaussian functions 
to fit these lines and to determine line centroids, EWs, and errors
(Table~\ref{tab:linedetection}). 
We found that all these lines are unresolved, i.e., the fitted line
widths are small enough to be identical to zero broadening, 
except for the apparently broad \oi\ K$\alpha$, which is caused by line 
saturation. For our spectral analysis purpose 
(see \S~\ref{sec:results}), we also used the Gaussian profiles
to measure the 95\% upper limits to the EWs of several nondetection lines
of our interest by fixing their centroids at the corresponding rest frame 
wavelengths (Table~\ref{tab:linedetection}). Figures~\ref{fig:counts} 
and \ref{fig:ratio} show these lines and edges in the count 
and the flux-normalized spectra, respectively. 

\begin{small}
\begin{deluxetable}{llcrcc}
  \tablewidth{0pt}
  \tablecaption{Line Measurements \label{tab:linedetection}}
  \tablehead{
& &    $\lambda$ & S/N  & EW \\
Ion & Transition &    (\AA) & ($\sigma$) & (m\AA) & log[N(cm$^{-2}$)]} 
  \startdata
\oi\     & 1s--2p  & $23.508^{+1.6}_{-1.6}   $ & 16.6   & $50.4^{+3.0}_{-2.9}$ & $17.85^{+0.05}_{-0.05}$ \\
\oi\     & 1s--3p  & 22.886(f)              &  $<1$  & $<16.1$                  & --- \\                   
\oii\    & 1s--2p  & $23.348^{+4.2}_{-4.2}   $ &  4.6   & $28.1^{+6.2}_{-6.2}$  & $17.01^{+0.21}_{-0.29}$ \\
\nei\    & 1s--3p  & $14.294^{+1.5}_{-1.3}   $ &  6.6   & $2.0^{+0.3}_{-0.3}$  & $17.14^{+0.06}_{-0.06}$ \\
\neii\   & 1s--2p  & $14.605^{+1.0}_{-1.0}   $ &  13.2  & $4.0^{+0.3}_{-0.3}$  & $16.71^{+0.04}_{-0.04}$ \\
\neii\   & 1s--3p  & $14.001^{+2.0}_{-1.2}   $ &  7.2   & $2.1^{+0.3}_{-0.3}$  & ---                     \\
\neiii\  & 1s--2p  & $14.507^{+2.0}_{-2.1}   $ &  6.8   & $2.1^{+0.3}_{-0.3}$  & $16.06^{+0.06}_{-0.07}$\\ 
\neiii\  & 1s--3p  & $13.690^{+6.3}_{-1.5}   $ &  3.3   & $0.9^{+0.3}_{-0.3}$  & ---                   \\  
\hline
\ovi\    & 1s--2p  & $22.026^{+4.0}_{-4.0}   $ &  4.4   & $10.8^{+2.4}_{-2.4}$& $15.73^{+0.17}_{-0.13}$  \\
\ovii\   & 1s--2p  & 21.602(f)              &  $<1$     & $<6.3$                 & $<15.35$ \\             
\ovii\   & 1s--3p  & $18.625^{+2.6}_{-2.5}   $ &  5.4   & $4.4^{+0.8}_{-0.8}$ & $16.06^{+0.08}_{-0.09}$\\  
\ovii\   & 1s--4p  & 17.765(f)	             &  $<2$	 & $<2.2$  & --- \\                                
\oviii\  & 1s--2p  & $18.964^{+2.0}_{-1.7}   $ &  6.1   & $5.1^{+0.8}_{-0.8}$ & $15.87^{+0.08}_{-0.08}$ \\ 
\oviii\  & 1s--3p  & $16.003^{+6.7}_{-6.7}   $ &  1.3   & $0.6^{+0.5}_{-0.5}$ &                      ---\\ 
\neviii\ & 1s--2p  & 13.660(f)               & $<1$     & $<0.58$             & $<14.77$                \\ 
\neix\   & 1s--2p  & $13.445^{+1.1}_{-1.2}   $ &  11.2  & $2.7^{+0.2}_{-0.2}$ & $15.45^{+0.05}_{-0.04}$ \\ 
\neix\   & 1s--3p  & $11.549^{+1.4}_{-3.4}   $ &  3.9   & $0.7^{+0.2}_{-0.2}$ & ---                     \\ 
\nex\    & 1s--2p  & 12.134(f)                 & $<1$   & $<0.09$             & $<14.23$                \\ 
\mgxi\   & 1s--2p  & $ 9.170^{+0.6}_{-1.2}   $ &  3.5   & $ 0.4^{+0.1}_{-0.1}$& $15.02^{+0.08}_{-0.09}$  \\
\fexvii\ & 2p--3d  & 15.010(f)              &  $<2$  & $<1.0$                 & < 14.32                  \\
\hline
??	 & 	   & $16.081^{+6.2}_{-5.7}   $ &  1.9   & $0.8^{+0.6}_{-0.6}$                            \\
??       &         & $13.930^{+4.8}_{-4.8}   $ &  2.1   & $0.6^{+0.3}_{-0.3}$                              
\enddata
\tablecomments{Wavelength uncertainties are given at 1$\sigma$ levels and in 
	units of m\AA. Equivalent widths (EWs) and uncertainties are in units 
	of m\AA. The $f$ in parentheses indicates nondetection lines, and 
	the 95\% upper limits to their EWs are calculated by fixing the line 
	centroid at the given wavelength. The column densities of ions are 
	obtained by fitting those lined with absorption line model 
	(\S~\ref{sec:results}). See text for the detail.}
\end{deluxetable}
\end{small}

\section{Absorption feature analysis and Results}
\label{sec:results}

In this section, we assume that the detected absorption edges and multiple 
absorption lines are produced in the ISM (see \S~\ref{sec:origin} for further 
discussion) and use them to probe the chemical and thermal properties of the 
ISM in various phases. In the following course we refer to the neutral and 
warm ionized ISM as the cool phase, 
which is traced by the transitions of neutral and mildly ionized ions.
To infer the more fundamental properties (e.g., ionic column density $N_i$ 
and the dispersion velocity $v_b$) of the intervening gas, we replace the
Gaussian profiles with the multiple {\sl absline} models (\S~\ref{sec:atomic})
in modeling the transitions listed in (Table~\ref{tab:linedetection}).
The profile of an absorption line is determined by 
$N_i$ and $v_b$ in addition to the three intrinsic parameters
$\lambda$, $f_{ij}$, and $\gamma$ (\S~\ref{sec:atomic}). Therefore
to obtain $N_i$ and its error for each ion, we always constrain
the $v_b$ value as the first step in the following analysis.

\subsection{Cool phase interstellar medium}
\label{sec:CISM}

Before discussing the properties of the cool phase of the ISM, we need to 
revisit the hydrogen properties along the Cyg~X-2 sight line.
\citet{tak02} obtained a total hydrogen column density 
by assembling the various forms including \HI, \hii, and H$_2$ based on
the \HI\ 21 cm, H$\alpha$, and CO emission observations, respectively.
In their \HI\ estimate, they used the old map of \citet{dic90}, which
offers pointing observations with a grid of $\sim48'$. We here
follow their steps but use the Leiden Argentine Bonn (LAB) 
Galactic \HI\ Survey \citep{kal05} that offers a $\sim30'$ spatial grid and 
1.3 ${\rm km~s^{-1}}$ velocity resolution. To infer the \HI\ column
density toward Cyg~X-2, we average the column densities from the four 
adjacent \HI\ observations with respect to 
their angular separation to Cyg~X-2 sight line,
\begin{equation}
N{\rm_{CygX2}} = \frac{\sum_{i=1}^4 N_i/d_i^2}{\sum_{i=1}^4 1/d_i^2}, 
\end{equation}
where $N_i$ is the column density along each individual sight line, and
$d_i$ is the angular separation between the sight line and Cyg~X-2.
Similarly, we define the corresponding 
uncertainty of this calculation as, 
\begin{equation} \Delta N{\rm_{CygX2}} = \frac{\sum_{i=1}^4 \Delta N_i/d_i^2}{\sum_{i=1}^4 1/d_i^2},\ {\rm and}\ \Delta N_i = |N_{\rm CygX2} - N_i|. \end{equation}
We obtain $N_{\rm CygX2}=1.85\times10^{21}~{\rm cm^{-2}}$
and $\Delta N_{\rm CygX2}=0.05\times10^{21}~{\rm cm^{-2}}$,
which is slightly smaller than $2.17\pm0.20\times10^{21}~{\rm cm^{-2}}$
obtained by \citet{tak02}. 

We use the new filling factor ($f$)  
for the warm ionized gas to estimate the \hii\
column density. In \citet{tak02}, the greatest uncertainty of the \hii\ 
estimate came from the unknown filling factor. 
By jointly analyzing the H$\alpha$ emission measures and the
pulsar dispersion measures toward 157 sight lines, 
\citet{ber06} found that the distributions of the warm ionized gas and 
its filling factor can be characterized as
\begin{equation} n(z) = n_0 e^{-z/n_h}, ~f(z)=f_0 e^{z/f_h}, \end{equation}
with the mid-plane and the scale height values 
$n_0=0.41\pm0.06~{\rm cm^{-3}}$, $f_0=0.05\pm0.01$, 
$n_h=710^{+180}_{-120}$ pc, and $f_h=670^{+200}_{-130}$ pc.
Along the Cyg~X-2 sight line, the H$\alpha$ EM and the \hii\ column 
density can then be written as
\begin{equation} EM = \frac{n_0^2}{\sin b} \int^\infty_0 f e^{-2z/z_h} dz\end{equation}
and
\begin{equation} N_{\rm HII} = \frac{n_0}{\sin b} \int_0^{D \sin b} f e^{-z/z_h} dz,\end{equation}
where $b$ and $D$ are the Galactic latitude and the distance of Cyg~X-2,
respectively. We explore the allowed parameter spaces of 
$n_0$, $n_h$, $f_0$, and $f_h$ that satisfy the EM toward the Cyg~X-2 sight 
line of $1.28\pm0.33\times10^{20}~{\rm cm^{-5}}$ (inferred from the 
H$\alpha$ map after a correction for the extinction; Takei \etal 2002),
and then use the allowed values to estimate the \hii\ to
$N_{\rm HII}=0.53_{-0.10}^{+0.23}\times10^{21}~{\rm cm^{-2}}$. 

Adopting the estimated H$_2$ value 
of $<0.10\times10^{21}~{\rm cm^{-2}}$ from \citet{tak02}, 
we obtain the total hydrogen.
Table~\ref{tab:hydrogen} compares the various components of hydrogen 
derived in this work and obtained by the Takei et al..

\begin{deluxetable}{lcc}
  \tablewidth{0pt}
  \tablecaption{Hydrogen column density toward Cyg~X-2}
  \tablehead{
 	& \multicolumn{2}{c}{Column Density} \\
        & \multicolumn{2}{c}{($10^{21}~{\rm cm^{-2}}$)} \\
Form	& Takei et al.   & This work }
  \startdata
\HI 	& $2.17\pm0.20$ & $1.85^{+0.05}_{-0.05}$ \\
\hii  	& $0.70\pm0.18^a$ & $0.53^{+0.22}_{-0.08}$\\
H$_2$ 	& $<0.10$       & ---\\
total   & $2.87^{+0.25}_{-0.23}$ & $2.38^{+0.23}_{-0.10}$
\enddata
\label{tab:hydrogen} 
\tablecomments{$^a$ Value derived with the assumed filling factor
	$f=0.2$ \citep{tak02}.}
\end{deluxetable}

We now measure the neutral column densities of Ne, O, and Mg
from their absorption edges. 
Since the 1s--3p absorption lines of \nei\ and \neii\ are well separated in 
the spectrum and the observed neon absorption edge is very close to 
the \nei\ 1s--3p transition (Fig.~\ref{fig:counts}; 
Tables~\ref{tab:absedge} and \ref{tab:linedetection}), 
it is therefore justified to believe that this edge is 
solely due to \nei\ (see also \citealt{jue06}). Adopting the absorption cross 
sections from \citet{bal92} (Table~\ref{tab:absedge}), we obtain 
log[N${\rm _{NeI}}$(cm$^{-2}$)] = 17.25(17.18, 17.31) at 90\% confidence level.
Similarly, we obtain log[N${\rm _{OI}}$(cm$^{-2}$)] = 17.84(17.78, 17.89).
For Mg, because of its low ionization energy 
(7.65 eV;  Moore 1970), the \mgii\ is believed to be the dominant ion 
in neutral hydrogen region among all charge states (e.g., \citealt{sav96}).
Therefore the edge measurement in fact yields the \mgii\ column density,
i.e., log[N${\rm _{MgII}}$(cm$^{-2}$)] = 16.92(16.77, 17.02).

We then use the absorption line model, {\sl absline} 
(\S~\ref{sec:atomic}; Yao \& Wang 2005), to infer the 
column densities ($N_i$) of each low ionization ions.
For \nei, the 1s--3p absorption line and the neutral absorption edge 
in fact trace the same gas. The same is true for \oi\ absorption edge 
and its 1s--2p and 1s--3p absorption transitions and other ions with 
multiple transitions (Table~\ref{tab:linedetection}). To assure the \nei\ 
and \oi\ column densities measured from the edges being consistent with those 
measured from the lines at 90\% confidence levels, we find that $v_b$ is 
allowed to vary between 0 and $75~{\rm km~s^{-1}}$. We use this velocity 
range in {\sl absline} models to infer column densities of all low ionization 
ions (Table~\ref{tab:linedetection}). During these measurements, the column 
density and the $v_b$ of a single ion in multiple transitions are linked. 

With these column densities, we can infer the chemical abundances of the
cool phase ISM. Neon is a noble element and is unlikely to be depleted into
dust grains. We then add up the \nei, \neii, and \neiii\ and compare the total
with the $N_H$. For oxygen, since \oi\ has a very similar ionization energy as 
\HI, the ratio $N_{\rm {OI}}/N_{\rm{HI}}$ thus approximates its abundance in 
the gas phase. Note that the ratio of $N_{\rm {OI}}/N_{\rm{HI}}$ is 
inconsistent with that of $N_{\rm {OII}}/N_{\rm{HII}}$. 
This discrepancy may reflect an either incomplete assessment of the true 
amount of \oii\ and/or \hii, or that the two ratios differ by nature.
In the case of the determination of the amount of ionized oxygen, there are 
a few issues to consider. While the column-density ratio of \oii/\oi\ obtained
in this work is consistent with that obtained by \citet{jue04} in a 
systematic survey of the complex oxygen absorption edge structure,
our search grid does not pick up significant amounts
of oxygen in higher ionization stages such as \oiii\ as in \citet{jue04}.
Including these amounts would slightly ease the discrepancy. The instrumental
resonance at the \oii\ 1s-2p location is generally well determined. On the 
other hand, any contributions to this resonance that might have been over
estimated would actually increase the discrepancy.
For magnesium, we simply compare the column
density of the dominant ion \mgii\ to that of the total hydrogen.
The abundances of Ne, O, and Mg in the cool ISM phase 
are reported in Table~\ref{tab:abund}.

\begin{deluxetable}{lcccccc}
  \tablewidth{0pt}
  \tablecaption{Number ratios of O, Ne, Mg, and Fe to H}
  \tablehead{
    &  X/H    & X/H  &X/H & X/H  & X/H & X/H    \\
X   & ($10^{-4}$) & ($10^{-4}$) & ($10^{-4}$) & ($10^{-4}$)& ($10^{-4}$) & ($10^{-4}$)}  
  \startdata
Ne & 1.23 & 1.00  & 0.69 & 0.87 & $0.84^{+0.13}_{-0.10}$  & ---	\\ 
O  & 8.51 & 5.45  & 4.57 & 4.90 & $3.83^{+0.48}_{-0.43}$  & $5.81^{+1.30}_{-1.34}$ \\ 
Mg & 0.39 & 0.35  & 0.34 & 0.25 & $0.35^{+0.09}_{-0.11}$  & $0.33^{+0.09}_{-0.09}$ \\
Fe & 0.32 & 0.28  & 0.28 & 0.27 & $\cdots$		  & $<0.31$ \\
\hline
Ref.& 1   & 2     & 3    & 4   & 5			  & 6
\enddata
\label{tab:abund} 
\tablecomments{References.---(1) \citet{and89}; 
	(2) \citet{hol01};
	(3) \citet{asp05};
	(4) \citet{wil00};
	(5) Abundances in the cool phase ISM derived in this work 
		(\S~\ref{sec:CISM}).
	(6) Abundances in the hot phase ISM derived in this work 
	(\S\S~\ref{sec:HISM} and \ref{sec:chemical}; Table~\ref{tab:absline}).
}
\end{deluxetable}

\subsection{Hot phase interstellar medium}
\label{sec:HISM}

As for the low ionization ions, we use multiple {\sl absline} 
models to infer column densities of highly ionized ions. 
Since in the hot ISM, the line broadening is dominated by the non-thermal 
velocity (e.g., turbulence; Savage \etal 2003; Yao \& Wang 2005, 2006),
we ignore the subtle thermal velocity difference of different ions 
and use the same $v_b$ value in the absline model for all transitions.
In this modeling, we also link the column density of individual ions 
in multiple transitions (e.g., 1s--2p and 1s--3p of \oviii; 
Table~\ref{tab:linedetection}), except for
the \ovii\ K$\alpha$ that has not been detected while the \ovii\ K$\beta$ 
line is significant (see \S~\ref{sec:oviimis} for further discussion).
By doing this, we find the $v_b$ can be directly constrained as
$104^{+43}_{-54}~{\rm km~s^{-1}}$ although any individual lines are 
unresolved (\S~\ref{sec:observation}). 
This constraint comes partly from the ratio of multiple transitions of same 
ions (e.g., Yao \& Wang 2006) and partly from using the same $v_b$ in all 
transitions. The latter is equivalent to co-adding all the detected absorption 
lines to obtain an S/N ratio of $>200$ per 10-m\AA\ spectral bin around an
absorption line. With the constrained $v_b$, we further infer 
the $N_i$ for different ions (Table~\ref{tab:linedetection}). 

We now use $N_i$ to probe the thermal and chemical properties of the hot ISM.
Since a significant \ovii~K$\alpha$ line has not been detected, it is not
used as a constraint in the following absorption line diagnostics.
Neither is \ovi\ used because it likely arises from a different location
from the $\sim10^6$ K gas discussed here 
(see \S\S~\ref{sec:halo} and \ref{sec:ovi} for further discussion).
If the hot gas is
isothermal, then the ratio of $N_i$ between the different
ionization states of the same ion (e.g., \ovii\ and \oviii) provides a unique
probe of its temperature and the equivalent hydrogen column density in the hot 
phase (e.g., \citealt{sut93, yao05}).
With the constrained temperature, we can further infer the relative abundances
of O/Ne, Mg/Ne, and Fe/Ne by comparing the $N_i$ of \ovii\ (or \oviii), \mgxi, 
and the nondetected \fexvii\ to that of \neix\ \citep{yao05}. We realize this 
powerful diagnostic by jointly analyzing the multiple high ionization 
absorption lines step by step, as illustrated in Table~\ref{tab:absline} 
(Case I).  Note that the above characterization of the hot gas only 
accounts for $<10\%$ of the observed \ovi\ absorptions.

\begin{deluxetable*}{llccccc}
  \tablewidth{0pt}
  \tablecaption{Step-by-step absorption line diagnostic of 
    the hot gas properties \label{tab:absline}}
  \small
  \tablehead{ 
   Included lines  & log[$N{\rm _H(cm^{-2})}$] & log[$T$(K)]  & $\Gamma$ & O/Ne & Mg/Ne & Fe/Ne } 
  \startdata
\sidehead{Case I}
   \ovii, \oviii\
	& $19.39^{+0.09}_{-0.07}$ & $6.28^{+0.03}_{-0.02}$ & $\cdots$ & $\cdots$ & $\cdots$ & $\cdots$ \\
   \ovii, \oviii, \neix 
	& --- & --- & $\cdots$ & $0.86^{+0.16}_{-0.15}$ & $\cdots$ & $\cdots$ \\
   \ovii, \oviii, \neix, \mgxi
 	& --- & --- & $\cdots$ & --- & $1.08^{+0.26}_{-0.27}$ & $\cdots$ \\
   \ovii, \oviii, \neix, \mgxi, \fexvii 
 	& --- & --- & $\cdots$ & --- & --- & $<1.4$ \\
\hline
\sidehead{Case II}
   \ovii, \oviii, \neviii, \neix, \nex, \mgxi, \fexvii
	& $19.38^{+0.05}_{-0.04}$ & $6.31^{+0.14}_{-0.04}$ & 10(1, 18) & 
	$1.00^{+0.19}_{-0.17}$ & $1.23^{+0.29}_{-0.26}$ & $<1.23$ 
    \enddata
  \tablecomments{The X/Ne ratio is in units of the solar value 
	of \citet{and89}. See text for the detail.}
\end{deluxetable*}

The derived chemical properties are robust against different characterizations
of the thermal properties of the hot ISM. The hot gas can extend as far as 
several kpc from the Galactic plane (\S~\ref{sec:intro} and references 
therein), therefore the isothermal assumption adopted above may not be 
sufficient to describe its thermal properties (e.g., \citealt{yao07, yao09}).
To examine the dependency of the obtained chemical abundances on 
different approximations to the gas thermal properties, we 
explore a more complex scenario by assuming that the column 
density of the hot gas follows a power law function of the gas temperature,
\begin{equation} N = N_0 \left(\frac{T}{T_{\rm max}}\right)^\Gamma. \end{equation}
Such a characterization can be derived assuming that both the gas density
and temperature decrease exponentially as a function of distance away 
from the Galactic plane \citep{yao07, yao09}. We then use this distribution to
jointly analyze all the highly ionized absorption lines used above
at the same time. To obtain a better constraint on the model parameters, 
we also include in our analysis the nondetection of \neviii\ 
and \nex\ K$\alpha$ transitions. The new constrained O/Ne, Mg/Ne, and Fe/Ne 
(Case II in Table~\ref{tab:absline}) are consistent with those derived in the 
isothermal case. This characterization 
can account for $<45\%$ of the observed \ovi\ absorptions.

\section{discussion}
\label{sec:dis}

\subsection{Location of the absorption}
\label{sec:origin}

\begin{figure}
   \plotone{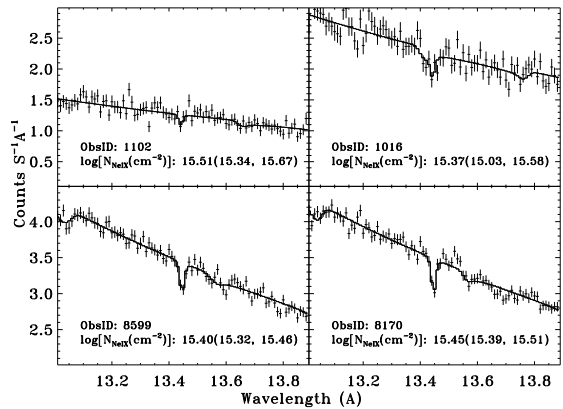}
   \caption{\neix\ absorption lines and the measured column densities
	at various source flux levels
	in the four different observations. The bin size is 10 m\AA.
    \label{fig:neix}
  }
\end{figure}

The intervening gas that is responsible for the observed absorption features
could be either the various phases of the ISM or the  
circumstellar material associated with the Cyg~X-2 system. 
Since the binary is moving toward the 
Sun at a velocity of $\sim220~{\rm km~s^{-1}}$ (\S~\ref{sec:intro}), 
the line kinematics are thus the most straightforward
way to distinguish these two scenarios. Unfortunately, the accuracy of the 
current atomic data make such a task very challenging.
For example, the newly calculated rest frame wavelength of \ovii\ K$\beta$ 
is off by $\sim400~{\rm km~s^{-1}}$ from the values 
commonly referenced (Table~\ref{tab:lineparameters} and references therein). 
The theoretical values for the low ionization ions are even less 
reliable, as already demonstrated in \citet{jue04, jue06}. To select more 
reliable rest frame wavelengths for high ionization ions, we picked up 
those values from Table~\ref{tab:lineparameters} that are consistent within 
$100~{\rm km~s^{-1}}$ among different references, which include \oviii, \neix, 
and \mgxi\ K$\alpha$ and \oviii\ K$\beta$ transitions. For low ionization ions,
we adopted the measured values from \citet{jue04, jue06},
which include \oi, \neii, and \neiii\ K$\alpha$ and \nei--\neiii\ K$\beta$.
We found that velocity shifts of our detected absorption lines are all
consistent with the reference ``rest frame wavelengths'' within 
$\sim100~{\rm km~s^{-1}}$. The statistical uncertainties of the line centroids
are usually much smaller due to the high spectral quality 
(Table~\ref{tab:linedetection}). On the other hand, if the absorption lines
reflect the circumstellar material at different ionization states, they are 
expected to be Doppler shifted and broadened at velocities comparable to the 
escape velocity of the system \citep{yao05}, which is inconsistent with the 
narrowness of the detected lines (\S~\ref{sec:results}). Furthermore, the 
consistent measurements of the \neix\ column density (Fig.~\ref{fig:neix}), 
for example, indicate a lack of ionization variation that could be 
expected at different source flux levels in a photoionized scenario.
The above arguments suggest that all the absorption
features are consistent with an ISM in origin, although the circumstellar 
scenario with some specific geometric effects cannot be completely ruled out.
It is worth noting that highly ionized circumstellar material has 
indeed been observed via broad \emph{emission} lines which are consistent with the
blueshift implied by the source's motion towards the Sun 
(\S~\ref{sec:oviimis} and \citealt{sch09}).

\subsection{References to the rest frame wavelengths}
\label{sec:restframe}

If the above detected absorption lines indeed arise from the multiple phases
of the ISM, their line centroids are a good reference for the rest 
frame wavelengths of the corresponding atomic transitions. Although the 
spectral resolution (the full width at half maximum; FWHM) of the MEG is
$\sim$23 m\AA, the centroid of a line can be measured as accurately as one 
tenth of that (e.g., \citealt{ish06} and Table~\ref{tab:linedetection}).
In contrast, theoretical calculations, especially for the low ionization
ions, are yet 
to merge (Table~\ref{tab:lineparameters}; \citealt{jue04, jue06}).

There are two main systematic uncertainties that affect the measurement of the
line centroids presented in this work. The first comes from the instrumental 
calibrations. As discussed in \S~\ref{sec:observation}, the first step in our 
data reduction is to determine the zeroth order source position; any 
uncertainty of the position will be linearly propagated to the whole 
wavelength scale. This uncertainty affects short wavelengths more
than it affects long wavelengths.
To quantify this effect, for each of the four observations 
(Table~\ref{tab:observation}), we compare the positive and the negative arms 
in measuring the \oi\ K$\alpha$ line, the longest wavelength line of our 
interest (Table~\ref{tab:linedetection}). 
We find that except for the short observation (ObsID 1016) in which there 
is not enough counts near the line, the
difference in the other three individual observations is less than 6.6 m\AA.
Since we measured the averaged value of the two arms 
in our final data analysis, we then take
the systematic error due to the calibration to be $<4.7$ 
($=6.6/\sqrt{3-1}$) m\AA.

The other source of the systematic bias is due to the differential rotation
of the ISM toward the Cyg~X-2 sight line. Gas in the Galactic disk at 
different radii is rotating at different velocities around 
the Galactic center (GC) in near circular orbits; gas close to the GC has
a shorter period than that further out. For gas at radius
$R$ and with a rotational velocity $V$ moves with respect to the local standard
rest frame (LSR) at a speed of
\begin{equation}
V_r = R_0\sin l \left( \frac{V}{R} - \frac{V_0}{R_0} \right), 
\end{equation}
where $R_0$ and $V_0$ are the radius of the LSR and its rotation velocity,
respectively \citep{spa00}. The observed differential velocity 
is then the integral of all $R$s and $V$s 
along the pathlength to Cyg~X-2.

The velocity shift due to the differential rotation is very small.
For low ionization ions, this effect can be revealed from the 21 cm \HI\ 
emission. Averaging the velocities with respect to the column densities for 
the four adjacent \HI\ observations around the Cyg~X-2 sight line 
(\S~\ref{sec:CISM}), we obtain $\overline{V_r} = -8~{\rm km~s^{-1}}$. 
For high ionization ions, since they extend to 
a much larger scale than the low ionization ones (\S~\ref{sec:intro}),
the ``halo-lagging'' effect (e.g., \citealt{rand97,rand00}) must be taken
into account. We here assume that, (1) the 
density of the hot gas decreases exponentially along the vertical distance away
from the Galactic plane with a scale height of 3 kpc (e.g., Bowen 
\etal 2008; Yao \etal 2008); (2) the rotation velocity linearly decreases from
$V_0$ and $V$ at the Galactic plane to zero at a height $z_0$ above the 
Galactic plane; and
(3) the velocity at radii between $R_0$ and $R$ can be linearly interpolated 
from $V_0$ and $V$. For the LSR, we take $R_0=8$ kpc and 
$V_0=220~{\rm km~s^{-1}}$, and for the Galactic radius of Cyg~X-2, $R=11$ kpc. 
Again, we average the $V_r$ with respect to the column density along the 
line of sight. For an extreme case, $z_0$ = 1.5 kpc (where Cyg~X-2 is 
located; \S~\ref{sec:intro}), we obtain an firm upper limit
of $\overline{V_r}<117~{\rm km~s^{-1}}$ by taking $V=164~{\rm km~s^{-1}}$  
that represents the smallest velocity derived from the smallest \HI\ emission 
velocity of $V_r=-100~{\rm km~s^{-1}}$ along the Cyg~X-2 sight line 
(\S~\ref{sec:CISM}). If assuming a more reasonable $z_0$ = 8 kpc, we then 
obtain 
$\overline{V_r}\sim56~{\rm km~s^{-1}}$, which corresponds to an uncertainty of
4.4, 4.1, and 1.7 m\AA\ at wavelength of 23.5, 22, and 9 \AA, respectively.

It is worth pointing out that besides the absorption lines presented in this
work, the highly ionized emission lines observed in stellar coronae
(e.g., \citealt{hue03}) also provide complementary references for the rest 
frame wavelengths of the atomic transitions. The wavelengths measured in this
work are remarkably consistent with those detected in stellar coronae.

For transitions of the low ionization ions, the absorption line centroids and 
the edge wavelengths reported in \citet{jue04, jue06}, though with larger
statistical errors due to the lower counting statistics in their used spectra,
are consistent with the values obtained in this work.  

\subsection{Mystery of the un-detected \ovii\ K$\alpha$ line at 21.6 \AA}
\label{sec:oviimis}

\begin{figure}
   \plotone{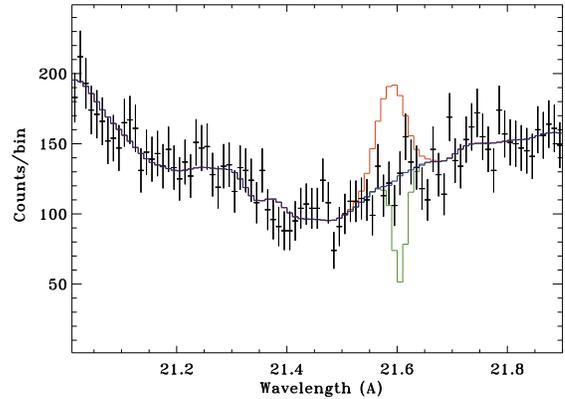}
   \caption{Observed spectrum, the best fit continuum (blue), and the
 	predicted \ovii\ K$\alpha$ line absorptions (green) from the observed 
	\ovii\ K$\beta$ transition (Table~\ref{tab:linedetection}).
	The red line marks the \ovii\ resonance emission needed to fill
	in the \ovii\ K$\alpha$ absorption.
    \label{fig:OVIIKa}
  }
\end{figure}

In our high quality spectrum, we did not detect the \ovii\ K$\alpha$
absorption line, which however, is expected to be one of the most
prominent lines because of the high abundance of oxygen, the high
ionization fraction of \ovii\ in a broad temperature range \citep{sut93}, 
and the large absorption oscillation strength of the transition
(Table~\ref{tab:lineparameters}).
Figure~\ref{fig:OVIIKa} exhibits the expected \ovii\ K$\alpha$ absorption
based on the measurement of the detected \ovii\ K$\beta$ 
and non-detected \ovii\ K$\gamma$ lines (Table~\ref{tab:linedetection}).

The absence of the \ovii\ K$\alpha$ absorption is puzzling. 
One obvious possibility would be that there
is an \ovii\ resonance emission line with a flux compensating for the
line absorption. If this is the case, it requires a line flux of 
$4.9\times10^{-4}~{\rm photons~s^{-1}~cm^{-2}}$. Emission
lines tracing the photoionized
accretion-disk-corona (ADC) have been observed in this 
source, but the lines are generally very broad with a minimum FWHM of 
$\sim800~{\rm km~s^{-1}}$ \citep{sch09}. Figure~\ref{fig:OVIIKa} demonstrates 
such a local broad emission line at the systematic velocity of 
Cyg~X-2 ($-220~{\rm km~s^{-1}}$; \S~\ref{sec:intro}) on top of the 
continuum; its flux is scaled to that of the missing \ovii\ K$\alpha$ at 
the corresponding wavelength. Although such a line has the power to 
compensate for the absorption, the slight offset as well as the stark 
difference in line broadening has to leave a significant residual
with respect to the line absorption. The ADC region is also expected to have a
very high density, and the intercombination line usually dominates 
the triplet emission for the He-like ions \citep{sch09}. If it
is the intercombination rather than the resonance transition that fills in
the absorption, besides its broadening, it is also expected to be blue-shifted
at $\sim2.7\times10^3~{\rm km~s^{-1}}$ for \ovii. Such a velocity shift was 
not observed in other detected emission lines \citep{sch09}.
Also, the intercombination line will be slightly broader than the resonance and
should leave a similar residual. The ionization parameters derived from the
emission lines are also too high to expect the emission from the oxygen ions
\citep{sch09}. 
Another source that could fill the expected \ovii\ K$\alpha$ 
line absorption is the scattered resonance line emission of the hot 
ISM. But this diffuse emission 
presumably uniformly fills the {\sl Chandra} field of view, and thus would only 
produce a very broad rather than a narrow emission enhancement in the 
grating spectrum. Futhermore, the intinsic \ovii\ intensity of 
the diffuse ISM along the Cyg~X-2 sight line is expected to be $\sim2.2-6.4$ 
${\rm photons~s^{-1}~cm^{-2}~str^{-1}}$ \citep{yos09}, and the foreground 
cool ISM will attenuate this emission to be 
$\sim0.3-0.9~{\rm photons~s^{-1}~cm^{-2}~str^{-1}}$ 
(taking $N_{\rm H} = 2.38\times10^{21}~{\rm cm^{-2}}$; 
Table~\ref{tab:hydrogen}) before the emission reaches detectors.
The small field of view ($8.3'\times16.6'$, the full field of view of two ACIS 
chips) of {\sl Chandra} thus could only yield a line flux 
of $0.03-0.1\times10^{-4}~{\rm photons~s^{-1}~cm^{-2}}$ at most, 
which is much less than the missing amount.
The consistent measurement of the non-detected \ovii~K$\alpha$ line between 
positive and negative grating spetral orders also rules out any possible 
instrumental effects (e.g., an unfortunate 
set of hot pixels or unrecognized non-X-ray events precisely at the position
of the absorption line on the detector) that could fill the absorption line.

The other detected absorption lines do not seem to be significantly 
diluted by any peculiar line emission. 
The putative ADC emission is not expected to affect our absorption 
measurements in general. The detected ISM lines are nearly unresolved  
(\S~\ref{sec:results}), 
and their measurements rely on the local continuum in a very narrow
wavelength range. The generally very broad emission lines as observed in the
Cyg~X-2 binary system can therefore be regarded as continuum and thus
will not generate any confusion to the
measurements presented in this work. The only concern with respect to our measurements
is a likelihood of the similar peculiar narrow emission lines as 
may possibly occur in the case of \ovii\ K$\alpha$. 
For the H- and He-like oxygen and neon ions, the emissivity ratios between
the 1s--2p and the 1s--3p transitions in the temperature range of
$10^5-10^7$ K are always larger than the ratios of the corresponding 
absorption oscillation strengths. Therefore any existing emission will fill 
more rapidly the 1s--2p lines than the 1s--3p ones.
We re-fit the individual K$\alpha$ and K$\beta$ absorption lines
of \oviii\ and \neix, and find that the column densities derived from the 
different transitions are consistent within 1$\sigma$ for each ion. 
In contrast, the 1s--3p of \ovii\ provides a very tight constraint on \ovii\ 
column density while the 1s--2p only yields an upper limit 
(Table~\ref{tab:linedetection}). This consistent measurement in 
\oviii\ and \neix\ could only be incidentally reached when the filled-up 
portions of the lines were proportional to their absorption 
oscillation strengths plus the saturation effects. We 
thus speculate that the chance for the existence of the peculiar 
emission lines for other ions is small. Limited by the scope of this work,
we prefer to leave the mystery of the missing \ovii\ K$\alpha$ absorption line 
unexplained, and assume that other \ovii\ transitions (e.g., K$\beta$ and 
K$\gamma$) still faithfully reflect the \ovii\ absorption column in the 
following sections.

\subsection{Chemical abundances in different phases}
\label{sec:chemical}

The chemical abundances of the cool phase ISM presented in 
Table~\ref{tab:abund} represent so far the best measurements of the 
absolute abundances in X-ray.
For a comparison, Table~\ref{tab:abund} also lists several references
for the solar/ISM abundances commonly used in the literature.
In these references, neon is the one of the most uncertain abundances
because its value is usually scaled with respect to other relatively 
well-studied ones like oxygen, due to the lack of suitable spectral lines in 
modeling the Sun's photosphere (e.g., \citealt{wil00, asp05}). 
In contrast, because the total hydrogen column density 
of the cool phase along the Cyg~X-2 sight line is available, the measured
neon value in this work 
offers a valuable reference for its abundance in the ISM.
Our measured oxygen abundance is systematically lower than all the
referenced values, indicating an oxygen depletion
into dust grains (see below), whereas the magnesium value suggests
a mild or no depletion (Table~\ref{tab:abund}).

For the hot phase ISM, since the hydrogen is expected to be fully ionized at 
temperatures $\sim10^6$ K, there is no usable atomic transition to measure its
column density. However, assuming that the neon abundance remains the same 
in both cool and hot phases, we can utilize the established neon abundance in 
the cool phase to derive the absolute abundances of oxygen, magnesium, and 
iron in the hot phase based on their measured relative abundances
(\S~\ref{sec:HISM}; Tables~\ref{tab:abund} and \ref{tab:absline}).

It is interesting to compare the abundance difference between the cool and hot
phases (Table~\ref{tab:abund}). The oxygen abundance in the hot phase 
is remarkably lower than the old solar value of \citet{and89}, but is
consistent with the recently revised solar value \citep{asp05}
and the expected value in the ISM \citep{wil00}. However, it is apparently
higher than that in the cool phase, which indicates dust grain sputtering 
and/or recent metal enrichment of the hot phase. 
If the dust grains were totally destroyed in the hot phase, the difference
indicates that $\sim$30\% of the total oxygen is likely 
depleted into dust grains in 
the cool phase. The consistent measurement of the magnesium abundance in 
both phases further confirms that very little or no magnesium is depleted
into dust grains.

If $\sim30\%$ ($N_{\rm O}=4.15\times10^{17}~{\rm cm^{-2}}$) 
of oxgyen is indeed depleted into dust grains, its 1s--2p transition 
that is analog to that of the atomic form at 23.51 \AA\
would also imprint a significant 
absorption feature near the position of the instrumental oxide resonance at
23.35 \AA. However, visual inspectation does not reveal any absorption line 
other than the one at 23.35 \AA\ between 22.81 \AA\ (the atomic oxygen 
edge position) and 23.51 \AA\ (Fig.~\ref{fig:counts}). Several previous works
actually attributed the absorption feature at 23.35 \AA\ to oxygen in
compound forms rather than to its atomic \oii\ 1s--2p transition
(e.g., \citealt{pae01, tak02}). However, such a attribution would further 
exacerbate the already deficit of the \oii\ column density 
(\S~\ref{sec:CISM}). One possible explanation is that the 1s--2p transition 
of oxygen in dust grains is completely suppressed due to the 
effectively filled L (n=2) shell and/or that the transition oscillation 
strength ($f_{\rm 12}$) for oxygen in compound forms is so small 
($<0.05$) that the absorption feature is too weak to be visible.

Let us compare these results with several previous works.

\citet{tak02} analyzed the {\sl Chandra} LETG observation of Cyg~X-2
(Table~\ref{tab:observation}) and measured the \oi\ and \neii\ 1s--2p
transitions and O and Ne absorption edges, but they attributed other features
in O K edge complex to compound features rather than 
to the atomic transitions (e.g., \oii\ and \oiii\ lines;
see \citealt{pae01, jue04}).
They also used multiple edge models to approximate the complex oxygen 
absorption edges due to the different compound compositions and different
ionization states. Interestingly, their overall measurements and derived
oxygen and neon abundances in the cool phases are consistent with ours
though with larger statistical errors.

\citet{jue04, jue06} systematically investigated the complex structures of the 
oxygen and neon absorption edges, and they also used a subset (ObsID 1102) of 
the data analyzed in this work (Table~\ref{tab:observation}).
Along the Cyg~X-2 sight line, the optical depth of the neon edge they obtained
is consistent with that we measured in this work, but their oxygen value is 
apparently higher. By examining 
the putative absorption features due to various compounds, they concluded that
their data allow 10\%-40\% of the oxygen to be depleted into dust grains, 
which is consistent with what we find above through comparing the O/Ne in 
cool and hot phases.

\citet{yao06} analyzed the multiple absorption lines in a {\sl Chandra}
spectrum of
4U~1820--303 in great detail and found that while about 50\% of the oxygen is 
depleted into dust grains in the cool phase, there is no evidence showing the
oxygen depletion in the hot phase, which in general agrees with 
what we find here. \citet{yao06b} further detected the \fexvii\ absorption
line in the {\sl Chandra} spectrum and concluded that a bulk of the heavily
depleted iron in the cool ISM as evidenced in both Far-UV and X-ray wavelength 
bands (e.g., Savage \& Sembach 1996; Juett \etal 2006) has been liberated 
during the dust sputtering process in the hot phase. However, in this work,
we did not detect the \fexvii\ line and only obtained an upper limit to the 
iron column density (Table~\ref{tab:linedetection}), which prevents us from
constraining a meaningful iron abundance 
(Tables~\ref{tab:abund} and~\ref{tab:absline}) and further testing the
iron depletion and dust grain destruction concluded in \citet{yao06}.
 
The small depletion of magnesium in the cool phase indicated in this work
contradicts the heavy ($\gsim70\%$) depletion previously concluded based on 
the measurement of the far-UV \mgii\ 
absorption lines in spectra of high latitude stars
(e.g., \citealt{sav96} and references therein). This discrepancy could be 
easily reconciled if a smaller absorption oscillation strength ($f_{ij}$) 
\footnote{The $f_{ij}$ values for \mgii\ lines at 
$\lambda\lambda$ 1240.4 and 1239.9 are still very uncertain.
For instance, $f_{ij}$ for $\lambda1239.9$ is $2.68\times10^{-4}$, 
$1.25\times10^{-3}$, and $6.32\times10^{-4}$ listed in \citet{mor91}, 
\citet{sav96}, and \citet{mor03}, respectively.}
is applied to these \mgii\ absorption lines.

\subsection{Limit to the large-scale Galactic Halo}
\label{sec:halo}

The detection of the high ionization absorption lines along the Cyg~X-2
sight line has implications for our understanding of galaxy formation 
and evolution. Highly ionized 
absorption lines (\ovii\ K$\alpha$ line in particular) 
have been commonly observed toward extragalactic sources 
(e.g., \citealt{fang06, bre07}). The absorptions could
arise from either the large-scale ($>20$ kpc) Galactic halo gas or the 
hot gas in the extended Galactic disk or a combination of both. 
The existence of the Galactic gaseous 
halo has been predicted in many semi-analytic calculations and numerical 
simulations of disk galaxy formation. The gas, originally accreted from
the intergalactic medium, could have been shock-heated to the virial
temperature during its in-falling into the dark matter halo's potential well
(e.g., \citealt{bir03}) and is believed to
contain the total baryonic mass comparable to or even greater than the total 
baryonic matter of stars and the ISM (e.g., \citealt{som06}). 
On the other hand, there is mounting
evidence showing the existence of the hot gas around the disk galaxies, 
for instance, the diffuse X-ray emission detected in disk 
galaxies (e.g., \citealt{tul06}), the spatial distribution of the 
far-UV \ovi\ absorption and emission (e.g., \citealt{sav03, dix08}), 
and the \ovii\ and \neix\ absorptions detected toward
the Galactic sources (e.g., \citealt{yao05, jue06}; this work). 
Both the morphology and the intensity of the diffuse X-ray emission in
the extragalactic sources suggest that the hot gas traces the stellar 
feedback in the galactic disk (e.g., \citealt{tul06}). The question
is, how much the disk hot gas contributes to the total X-ray absorptions
observed in the extragalactic sources. \citet{ras03} combined the 
absorption lines detected along three AGN (Mrk~421, 3C~273, and 
PKS~2155-304) sight lines with the \ovii\ emission measurement from
the sounding rocket experiment \citep{mcc02} and estimated a scale length of 
the absorbing/emitting gas to be $>140$ kpc. In contrast, \citet{yao07} 
extensively analyzed the absorption lines observed along the Mrk~421 sight line
and concluded that the absorbing gas is of a nonisothermal nature
and is consistent with a location of several kpc around the Galactic plane. 
The latter authors further argued that the discrepancy between their results
and those obtained by \citet{ras03} could be easily reconciled if the 
metallicity, density distribution, and nonisothermality of the 
absorbing/emitting gas were considered.

A more direct estimate of the Galactic disk contribution is to compare 
absorption lines observed toward Galactic sources with those observed toward
extragalactic sources.
\citet{yao08} recently conducted such a 
differential study by comparing the highly ionized
X-ray absorption lines observed towards three sight lines, a  
distant quasar Mrk~421, LMC~X-3 at a distance of $\sim50$ kpc 
in the Large Magellanic Cloud, and a Galactic sight line 4U~1957+11, 
and concluded that all X-ray absorptions toward the extragalactic sources 
can be attributed to the Galactic disk hot gas. 

A major uncertainty of the differential comparison performed by \citet{yao08}
is the possible non-homogeneity of the disk hot gas. 
Apparent variation of the X-ray absorptions has been detected 
along different AGN sight lines (e.g., \citealt{bre07}), which 
however, may not reflect the disk hot gas variation on a global scale
because nearly all the observed absorption enhancements 
are due to the known additional absorption components (e.g., the
north polar spur, the Galactic bulge region, etc.).
Both far-UV \ovi\ absorption and emission could vary
at an angular scale as small as $25'$ \citep{how02,dix06}. For a gas in the
CIE state, the population of \ovi\ peaks at intermediate temperatures at which 
the hot gas cools very efficiently; a bulk of the observed \ovi\ is believed
to exist at interfaces of the cool and hot gases (e.g., \citealt{sav06}). The 
\ovi\ thus traces more directly to the embedded cool gas clouds rather than to
the hot gas itself. The smooth distributions of the diffuse soft X-ray 
emission intensity in our Galaxy 
(again, except for the known emission enhancement regions like 
the north polar spur, the Galactic center, the Cygnus loop, etc.; 
\citealt{sno97}) and around the extragalactic 
disk galaxies \citep{tul06} in fact suggest that the galactic disk gas 
is likely to be homogeneous in general.

\begin{figure}
   \plotone{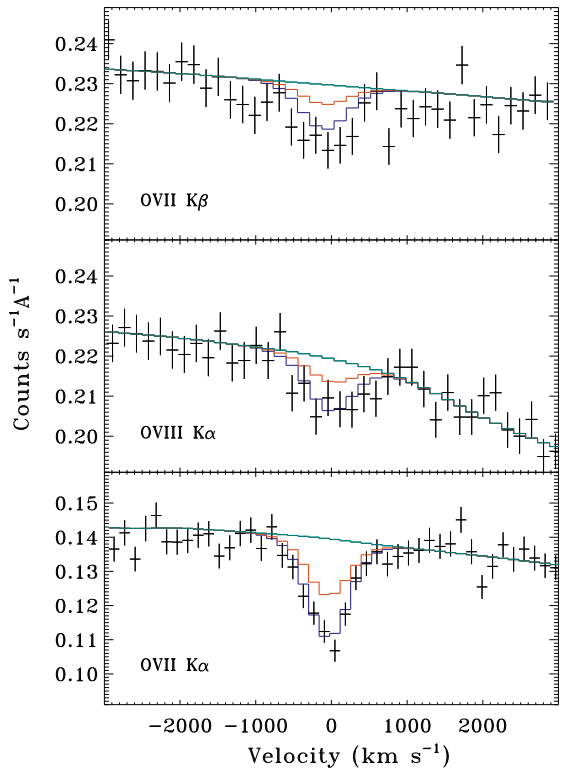}
   \caption{\ovii\ K$\beta$, \oviii\ K$\alpha$, and \ovii\ K$\alpha$ 
	absorption lines observed in the spectrum of Mrk~421 (cross),
	overlayed with the best fit continuum (green). The red histograms 
	mark the amount of the absorptions observed toward
	Cyg~X-2 sight line after the correction for the Galactic latitude
	dependence. The blue histograms indicate the total 
	expected Galactic disk absorptions by scaling the observed absorptions
	along the Cyg~X-2 pathlength with an exponential scale height of
	3 kpc. See text for the detail.
    \label{fig:mrkcyg}
  }
\end{figure}

We now compare the highly ionized absorption lines between the Cyg~X-2 sight
line and the extragalactic sight line Mrk~421. The latter presents to date the 
best constrained highly ionized absorptions toward extragalactic sight lines
(e.g., \citealt{yao07}). With respect to the Galactic X-ray binary 
4U~1957+11 (Galactic coordinates $l, b=51\fdg31, -9\fdg33$) used in
\citet{yao08}, Cyg~X-2 is located further away
from the Galactic center region and thus the observed absorption should be less
affected by the Galactic bulge gas.
We jointly analyze the absorption lines toward these two sources by assuming
the common disk hot gas has the same thermal properties. 
We normalize the absorption toward the Cyg~X-2 sight line to that toward
the Mrk~421 direction by assuming that the absorbing gas is slab-like 
distributed (e.g., by multiplying a factor of $\sin b$), and obtain the 
``net'' absorption beyond the Cyg~X-2 as 
$\log N{\rm_{OVII}(cm^{-2})} = 15.56(15.45, 15.64)$, and
$\log N{\rm_{OVII}(cm^{-2})} = 15.43(15.27, 15.55)$, i.e.,
the pathlength of Cyg~X-2 can account for $\sim40\%$ of the total absorption 
toward Mrk~421 sight line. If we further consider the location of Cyg~X-2
and assume that its pathlength only samples $<54\%$ of the disk hot gas 
in the vertical direction (taking its distance of $<12$ kpc and an 
exponential scale height of 3 kpc; \S~\ref{sec:intro} and \citealt{yao09}), 
we find that all the X-ray absorption toward 
the Mrk~421 can be attributed to the Galactic disk hot gas, and 
obtain upper limits to the ``net'' absorption beyond the disk as
$\log N{\rm_{OVII}(cm^{-2})}<15.32$ and 
$\log N{\rm_{OVIII}(cm^{-2})}<15.32$ (Fig.~\ref{fig:mrkcyg}).
This is consistent with the limits derived by \citet{yao08}.

\subsection{Predicted far-UV absorption and emission}
\label{sec:ovi}

\begin{figure}
   \plotone{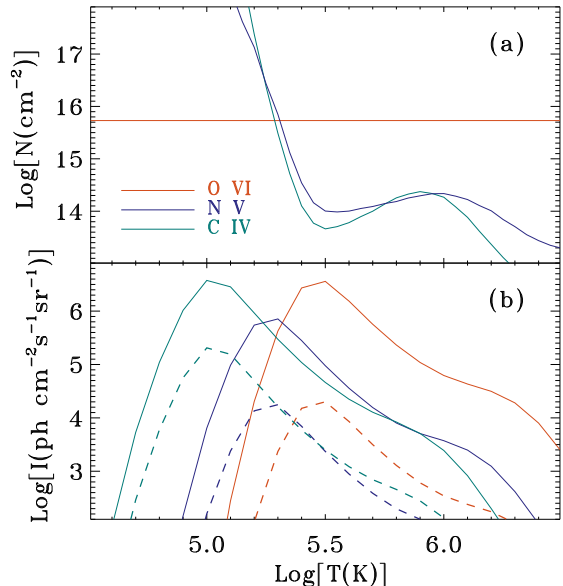}
   \caption{ (a) Detected \ovi\ and expected \nv\ and \civ\ column densities 
	as a function of temperature. 
	(b) Predicted doublet line emission intensity for He-like ions
	\ovi, \nv, and \civ\ as a function of temperature. 
	The dotted lines are the attenuated intensities 
	assuming all emission arising behind dust grains.
    \label{fig:intensi}
  }
\end{figure}

A large amount of far-UV absorption and emission is expected to arise from
the observed \ovi-bearing gas. The detected \ovi\ along the Cyg~X-2 sight line 
($\log N{\rm_{OVI}[cm^{-2}]}=15.73^{+0.17}_{-0.13}$;
Table~\ref{tab:linedetection}) is the largest \ovi\ column density observed 
so far. If assuming the solar abundance ratios of N/O and C/O, we can estimate
the expected \nv\ and \civ\ column densities (and thus the corresponding EWs 
of their L transition lines) associated with the \ovi-bearing gas for the 
isothermal assumption (\S~\ref{sec:HISM}).
Unlike the LMC~X-3 sight line toward which the hot \ovii-bearing gas can
explain all the observed \ovi\ absorptions \citep{yao09}, the hot 
($T\sim10^6$ K) gas toward 
the Cyg~X-2 sight line accounts for only a tiny portion of the observed 
\ovi\ (\S~\ref{sec:HISM}), leaving much of the \ovi\ to arise from the 
conductive interfaces (e.g., \citealt{sla89, sav06}).
\citet{dix06} recently derived an electron density of $n_e=0.29~{\rm cm^{-3}}$
at the interfaces
by comparing the \ovi\ emission and absorption along the same sight line. 
If taking this density and also assuming the solar abundance
ratios, we can estimate the 
expected \ovi, \nv, and \civ\ emission line intensities from the interfaces. 
Note that if assuming 55\% (\S~\ref{sec:HISM})
of the observed \ovi\ arise from the interfaces, 
the adopted electron density yields
an \ovi-gas size of 34 pc or a filling factor of 4\% for \ovi\ at its
peak ionization fraction.
Figure~\ref{fig:intensi} shows the expected \nv\ and \civ\ column
densities and the predicted intrinsic and attenuated 
line intensities by taking the reddening $E(b-V)=0.4$ \citep{mcc84} and the 
extinction from \citet{fit99}. These absorption and emission lines,  
\nv\ and \civ\ in particular,
are measurable and thus can be tested 
with the Cosmic Origins Spectrograph (COS) and the
Space Telescope Imaging Spectrograph (STIS) installed on
{\sl Hubble Space Telescope} (HST) with reasonably long exposures.

\section{Summary}
\label{sec:sum}

We presented a high resolution X-ray spectroscopy study of the 
multiphase interstellar medium based on four observations of Cyg~X-2 
with the {\sl Chandra} High Energy Transmission Grating Spectrometer. Our 
main results and conclusion can be summarized as follows:

1. We measure the properties of 14 identified absorption lines and 3 
absorption edges in the final co-added spectrum. 
Modeling these lines and edges we obtain the column 
densities of \oi-\oii, \nei-\neiii, \ovi-\oviii, \neix, and \mgxi.

2. We demonstrate that these absorption features trace the various phases of the 
interstellar medium rather than circumstellar material dynamically associated
with the Cyg~X-2 system. The well constrained centroids of the
absorption features provide a good reference for the rest frame wavelengths
of the corresponding atomic transitions.

3. Comparison between absorption edges and absroption lines of
\nei\ and \oi\ constrains velocity dispersion in cool phase 
ISM to be $v_{\rm b}<75~{\rm km~s^{-1}}$, and joint analysis of 
multiple high ionization absorption lines (\ovi-\oviii, \neix, and \mgxi)
yields $v_{\rm b}$ in hot phase ISM to be 
$104_{-54}^{+43}~{\rm km~s^{-1}}$.

4. Complementing the absorption data with \HI\ 21 cm, H$\alpha$, and CO 
emission data, we derive absolute abundances of neon, oxygen, and 
magnesium in the cool phase ISM to 
$0.89^{+0.13}_{-0.11}\times10^{-4}$, $3.83^{+0.48}_{-0.43}\times10^{-4}$,
and $0.35^{+0.09}_{-0.11}\times10^{-4}$, respectively. 
By jointly analyzing the multiple high ionized 
absorption lines we also derive the abundances of oxygen and magnesium
in the hot phase as $5.30^{+1.18}_{-1.33}\times10^{-4}$
and $0.31^{+0.08}_{-0.09}\times10^{-4}$, which are robust against different
assumed temperature distributions of the absorbing gas. In the cool phase, while
about 30\% of the oxygen is depleted into dust grains, there is no
evidence for magnesium depletion.

5. The observed \ovii\ and \oviii\ absorptions toward Cyg~X-2 can  
already account for $\sim40\%$ of the high ionization absorptions observed toward 
the Mrk~421 sight line. By considering the location of Cyg~X-2 and the
spatial distribution of the Galactic hot gas, this means that all the high ionization 
absorptions observed toward the extragalactic sight line can be attributed 
to the extended Galactic disk.

6. A large amount of far-UV absorption and emission is expected to arise 
from the \ovi-bearing gas. The expected \nv\ and \civ\ column densities
and the predicted emission line intensites are measurable with
the COS and STIS aboard {\sl HST}.

\acknowledgements 
We thank Q. Daniel Wang, J. Michael Shull, and Frits Paerels
for the extensive discussions 
on the detections and the non-detections of the absorption lines
presented in this work. We also thank an anonymous referee for 
insightful and constructive suggestions.
We are grateful to Thoms Gorzcyca for discussions on atomic data.
This work is supported by NASA through the 
Smithsonian Astrophysical Observatory contract SV3-73016 to 
MIT for support of the Chandra X-Ray Center under contract NAS 08-03060. 
YY also thanks for funding support from NASA grant    
NNX08AC14G, provided to the University of Colorado    
to support data analysis and scientific discoveries    
related to the Cosmic Origins Spectrograph on the 
Hubble Space Telescope.

\end{document}